\documentclass[%
 reprint,
 amsmath,amssymb,
 aps,
]{revtex4-1}

\usepackage{graphicx}
\usepackage{dcolumn}
\usepackage{bm}
\usepackage{mhchem}

\begin{document}


\title{Effects of Temperature and Near Ultraviolet Light on Current-Voltage Characteristics of Colemanite}

\author{Chandrima Chatterjee}
\email{cchatter@go.olemiss.edu}
 \affiliation{Department of Physics and Astronomy, University of Mississippi, University, Mississippi, USA 38677}
 
\author{Bhaskar Roy Bardhan}
 \email{bardhan@geneseo.edu}
\affiliation{Department of Physics and Astronomy, State University of New York, Geneseo, New York, USA 14454}

\date{\today}

\begin{abstract}
We investigate current-voltage (I-V) characteristics of the ferroelectric material colemanite $\textrm{Ca}_{2}\textrm{B}_{6}\textrm{O}_{11}.5\textrm{H}_{2}\textrm{O}$ at room temperature, at a high temperature of $400^{0}$F and under the influence of near ultraviolet light. We demonstrate that all three I-V plots exhibit hysteresis effects, and these new results shed new light on the resistance of colemanite. These novel properties are explained on the basis of its microstructure indicating potential applications in devices with negative resistance as well as in photovoltaic devices.
 \begin{description}

\item[PACS numbers]
61.72.-y, 77.80.-e, 77.84.-s, 91.60.Ed, 61.66.Fn, 77.84.Bw.
 
\end{description}
\end{abstract}

\pacs{Valid PACS appear here}
                              
\maketitle


\section{\label{sec:level1}INTRODUCTION}

Ferroelectric materials such as colemanite, have the ability to switch their spontaneous polarization by the application of an electric field. This property makes them suitable for applications in the electronics industry, medicine, and others~\cite{Pintilie}. In most of these applications, the ferroelectric materials are used as a capacitor, either as a bulk or single crystal, or as a polycrystalline thin film. In most of these applications, an external field is applied to the ferroelectric capacitor leading to an undesired leakage current. At a higher temperature, the pyroelectric effects may be dominant in materials such as colemanite, leading to a high leakage. As a result, any contributions from polarization switching would be suppressed. It is therefore extremely important to study the I-V characteristics of colemanite to explain the conduction mechanisms and its effect on macro-properties such as pyroelectricity.

In spite of the potential of colemanite to be used in the applications specified above, its ferroelectric, piezoelectric and pyroelectric properties have not yet been extensively investigated. Colemanite, whose chemical formula is \ce{Ca_2B_6O_11.5H_2O}, is a borate mineral which belongs to the class of prismatic monoclinic crystals. 	It is a hydrous compound and the impact of hydrogen bonding on crystal properties such as conductivity is not fully understood. The presence of hydrogen has been shown to affect the proton diffusion in perovskites such as Barium cerate~\cite{Swift}. In addition, colemanite contains both interstitial and substitutional defects which are introduced by the paraelectric-ferroelectric phase transition~\cite{Gavrilova}. The presence of defects have been found to affect the conductivity properties of several ferroelectric crystals such as Lithium niobate~\cite{bhatt2012,bhatt2017} and Lithium tantalate~\cite{Bhaumik}.


Ferroelectric materials, in general, are favorable materials for photovoltaic devices~\cite{Blouzon} due to the presence of an intrinsic electric field. However, they have high band gaps that make them unsuitable for many applications. At present, semiconductors are used in photovoltaic cells since they have low band gaps~\cite{Blouzon}, but the voltage generated is low. This has stimulated research in using ferroelectric materials as photovoltaic devices. The main challenges faced are unknown mechanisms of photoelectricity generation and device design~\cite{Inoue}. Most of the research in ferroelectric photovoltaics has been done using perovskites~\cite{Manser, Kojima, Snaith, Mohammed}. This has motivated us to investigate the photovoltaic response of other groups of ferroelectric materials such as colemanite where, to the best of our knowledge, no prior study has been conducted.

In this paper, we aim to investigate the effects of temperature on the dc current-voltage (I-V) characteristics of colemanite. The effects of near ultraviolet (UV) light on the I-V characteristics of colemanite are studied as well, and these results are explained based on the microstructure of the material.  In addition, we shed light on the photovoltaics of colemanite which provide a better understanding of photoelectricity generation with colemanite. This could potentially be widely applicable in low-budget energy harvesting.

This paper is organized as follows. In Sec. II, the structure of colemanite is discussed. This section includes the shape of the molecule, the arrangement of the host atoms and the possible defects present in colemanite, which are likely to be responsible for the ferroelectric properties of colemanite. In Sec III, the experimental techniques used to observe the I-V characteristics at room temperature, at $400^{0}$F and under excitation by near UV light are discussed. Section IV contains the experimental results  with discussions. Finally, a brief summary of the results with potential applications of colemanite are presented in Sec. V.

\section{\label{sec:level2}THE STRUCTURE OF COLEMANITE}
The basic structure of colemanite consists of a six membered ring of alternating B and O atoms as illustrated in Fig.~1. Two of the B atoms are tetrahedrally coordinated while the third one is triangularly coordinated. Each B atom is bonded with the O of a hydroxyl group. The six membered rings continue into infinite chains via shared O atoms. The chains are laterally linked to \ce{Ca^2+} ions which form sheets perpendicular to the six membered rings of B and O~\cite{Watton}.

\begin{figure}
\includegraphics [height=2.5in,width=3.5in]{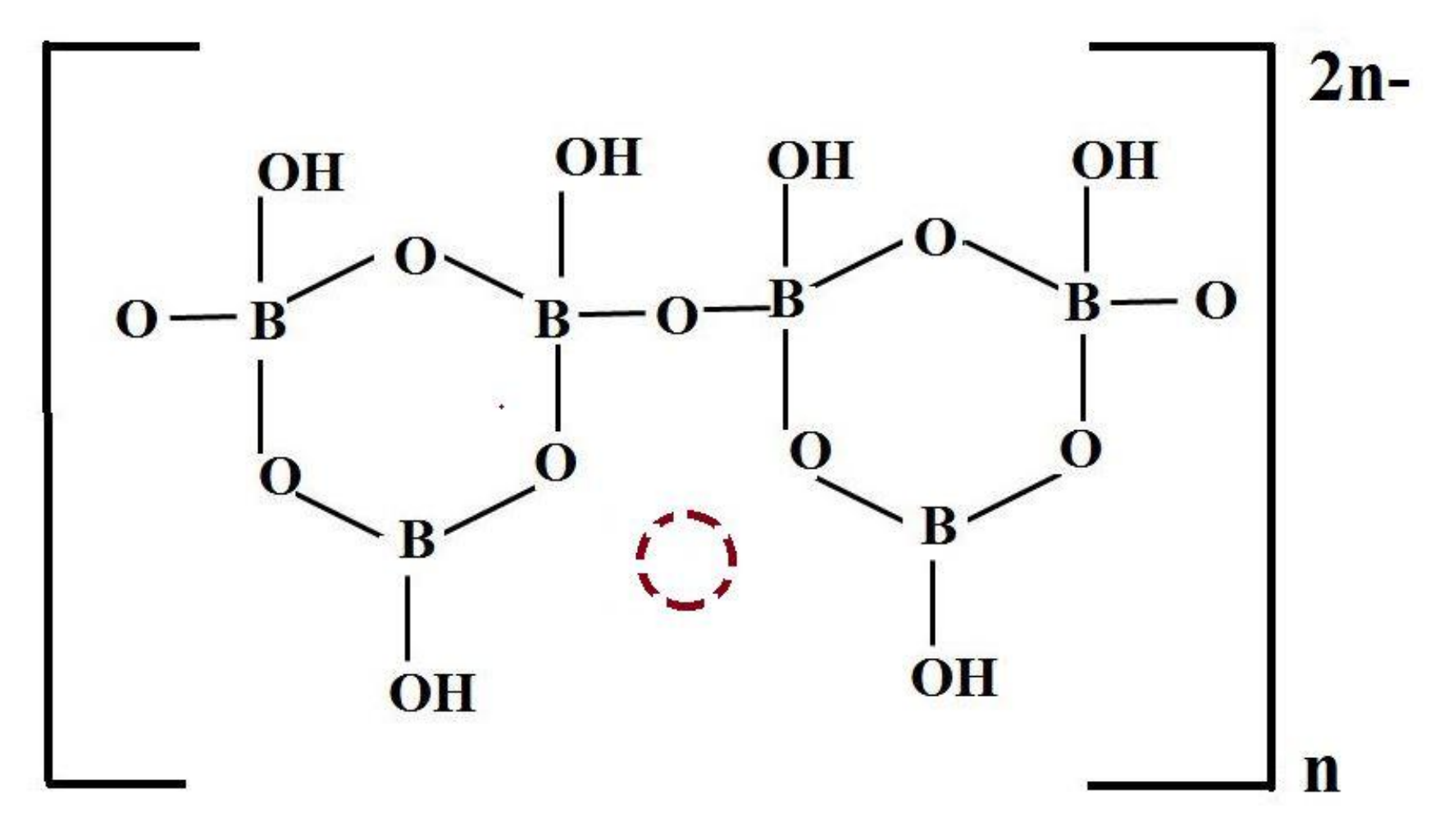}
\caption{Schematic representation of the structure of colemanite. The infinite polymerized chain is contained within the square bracket. The dashed circle represents the \ce{Ca^2+} ion which is found perpendicular to the direction of the polymerized chain.}
\end{figure}

Origin of ferroelectricity in colemanite can be explained from the structure shown in Fig.~1. The rings by themselves are non-polar, however the \ce{Ca^2-} and \ce{O^2-} coordination is polar. As a result, there occurs a spontaneous polarization in a direction perpendicular to the direction of the polymerized chain, which gives rise to ferroelectric domains.

Neutron diffraction studies have revealed that the ordering of H atoms in colemanite is temperature-dependent~\cite{Watton,Newnham}. At higher temperatures, the hydrogen atoms of the water molecule and a neighboring hydroxyl group, are in a state of disorder~\cite{Newnham}. As the temperature is lowered, the hydrogen settles into ordered non-centric positions in the ferroelectric phase. Colemanite contains a number of hydrogen bonds which are bifurcated or over-coordinated. The bifurcated hydrogen bonds are responsible for the anomalous behavior of water at \ce{4^oC}. The breaking and rearrangement of the bifurcated hydrogen bonds create quasi-free protons which may be responsible for conductivity of colemanite. The ordering of protons in hydrogen bonds are responsible for phase transitions in ferroelectrics such as Rochelle salt and other polymers whose structure contains water \cite{Gavrilova}.

 Unlike most ferroelectric materials, the domains in colemanite are not visible under plane polarized light. This is probably due to the presence of a mirror symmetry through the interconnecting O atom as seen in Fig. 1. Due to this enantiomer like property, two six membered rings on either side of the O atom, would rotate plane-polarized light by equal amounts in opposite directions, thus making it invisible. This would make colemanite a racemic compound and could be used in optical applications that require zero net rotation of plane-polarized light.  
 
Natural colemanite crystals are often mingled with crystallized gypsum \cite{Keyes} as the two tend to grow on the same bed of the earth. Hence, some of the common defects in colemanite would be Ca, S, O, and H  occurring as interstitial defects, substitutional defects and others. The paraelectric-ferroelectric phase transition in colemanite occurs over a wide range of temperature \cite{Gavrilova}. This would also create defects such as vacancies, color centers (or F-center), antisite defects and others. These defects are observed in ferroelectrics with perovskite structures such as Lithium niobate \cite{bhatt2012, bhatt2017} and Lithium tantalate \cite{Bhaumik}, which are subjected to heat treatment.

\section{\label{sec:level3}EXPERIMENTAL TECHNIQUES}
The sample used for the experiments consisted of a commercially available epitaxial colemanite crystal from Amargosa Minerals.  The dimensions of the sample were $1~\textrm{inch}\times0.9~\textrm{inch} \times 0.5~\textrm{inch}$.  In order to measure the I-V characteristics, copper electrodes were deposited on the front and back surfaces, with a layer of epoxy between the sample and the electrode. This helps in overcoming the Schottky barrier at the junction of the metallic electrode and the sample. A dc voltage was supplied to the sample using a regulated power supply (MW122A). The minimum voltage supplied by the instrument is 3V. The output current was observed with the help of a digital multimeter. In order to measure the I-V characteristics at a high temperature, the experimental setup was inserted in an oven. In order to uniformly heat the sample from all sides, the sample was positioned at the center of the oven. The measurements were carried out at a stable temperature of $400^{0}$F. At temperatures lower than this,  the output current was fluctuating over a wide range and a steady reading could not be taken.

In order to study the effect of near UV light on the I-V characteristics, the sample was excited with light at a wavelength of 395 nm, at the room temperature.The diameter of the light source was nearly twice larger than the sample. This ensured that the light was incident uniformly throughout the sample. The experimental setup was then moved to a dark chamber to avoid light from other sources. Under this condition, the I-V characteristics of colemanite were recorded.

\section{\label{sec:level3}RESULTS AND DISCUSSIONS}

\subsection{\label{sec:level2}I-V characteristics at room temperature}

Figure 2 shows the I-V characteristics of colemanite at the room temperature. The I-V plot shows hysteresis effects or path dependence the I-V characteristics. As the voltage is increased from 3 V, the current gradually increases until it reaches a maximum value of about 10 mA at 7 V. The crystal acts as an ohmic resistor up to this point. As the voltage is further increased, the current decreases until 10 V. After that, the current becomes constant. Thus, the crystal offers negative resistance from 7 V to 10 V. Due to the negative resistance, colemanite can be used as an amplifier. It can also be used to generate microwave energy which is produced with negative resistance devices \cite{golio}. As the voltage is now decreased, it is observed that the current decreases, until it reaches a minimum of about 1.5 mA at 6.5 V. Here the crystal acts as a ohmic resistor again. When the voltage is further reduced, the current almost increases in a straight line, thereby exhibiting negative resistance in this region.

\begin{figure}
\centering\includegraphics [width=3.5in,height=2.9in]{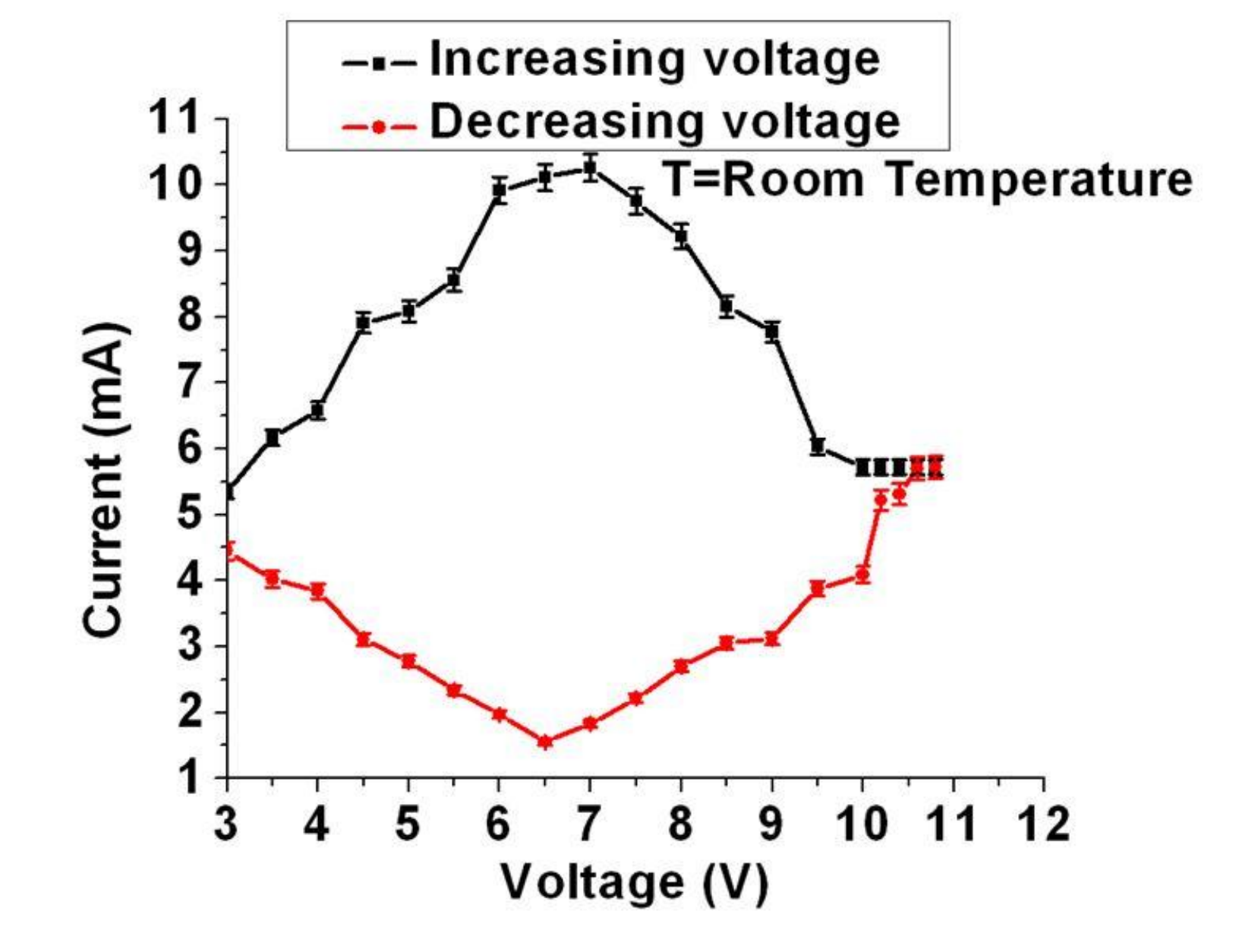}
\caption{I-V characteristics of colemanite at room temperature. When the instrument was turned on but the voltage was set to 0 V, the output current was 0.4 mA. This offset current is the source of error in the experiment.}
\end{figure}

We now explain, in details, the occurrence of such hysteresis effects in the I-V characteristics of colemanite based on its microstructure. As the voltage is increased, the internal dipole moments in a ferroelectric material align themselves in a single direction. This is the region beyond 10 V where the current is constant. As the direction of the voltage is reversed, the dipole moments gradually align in the opposite direction. The polarization switching is slow and takes place over a range of 7 V as observed in Fig. 2. The switching consists of three steps: the nucleation of ferroelectric domains with opposite direction of polarization, the growth of the domains with polarization parallel to the electric field, the compensation of the depolarization field occurring just after the switching takes place~\cite{Bhaumik}. The compensation is made by charges coming from the bulk of the material or from the electrodes. This step determines the speed of switching. If the ferroelectric contains plenty of free charges, the switching takes place very fast, over a couple of millivolts, and the hysteresis loop is expected to be rectangular. That is not the case for colemanite. The large range of voltage for switching implies a long time for compensating the depolarizing field. This can be explained by the presence of a large number of crystal defects which act as traps for free carriers. Some of these defects would be F-centers which are vacant lattice sites trapping electrons, antisite defects trapping electrons and mobile charged ions, grain boundaries trapping mobile carriers, and others. The result is a much higher resistivity which shows up as negative resistance at higher voltages. However, at low voltages, the defects act as low resistance conduction paths between the electrodes~\cite{Bhaumik}, leading to large currents. As the voltage increases, the paths start breaking down. The high conduction paths gradually disappear leading to a decrease of current. 

Decreasing the voltage in the other direction, decreases the energy available for the electrons to break free from the traps and start conducting. The current decreases until a certain threshold voltage of 6.5 V. Reducing the voltage beyond that re-establishes the conducting paths between the electrodes which increases current.
It can be observed that the range of voltages between which colemanite acts as an ohmic resistor in the forward direction, is the same range between which the crystal exhibits negative resistance in the backward direction, and vice-versa. In other words, the peak and valley of the plot are found in the same range of voltage. Thus the transition from offering a positive to negative resistance and vice-versa in both directions happens at nearly the same voltage.

\begin{figure}
\centering\includegraphics[width=3.5in,height=2.7in]{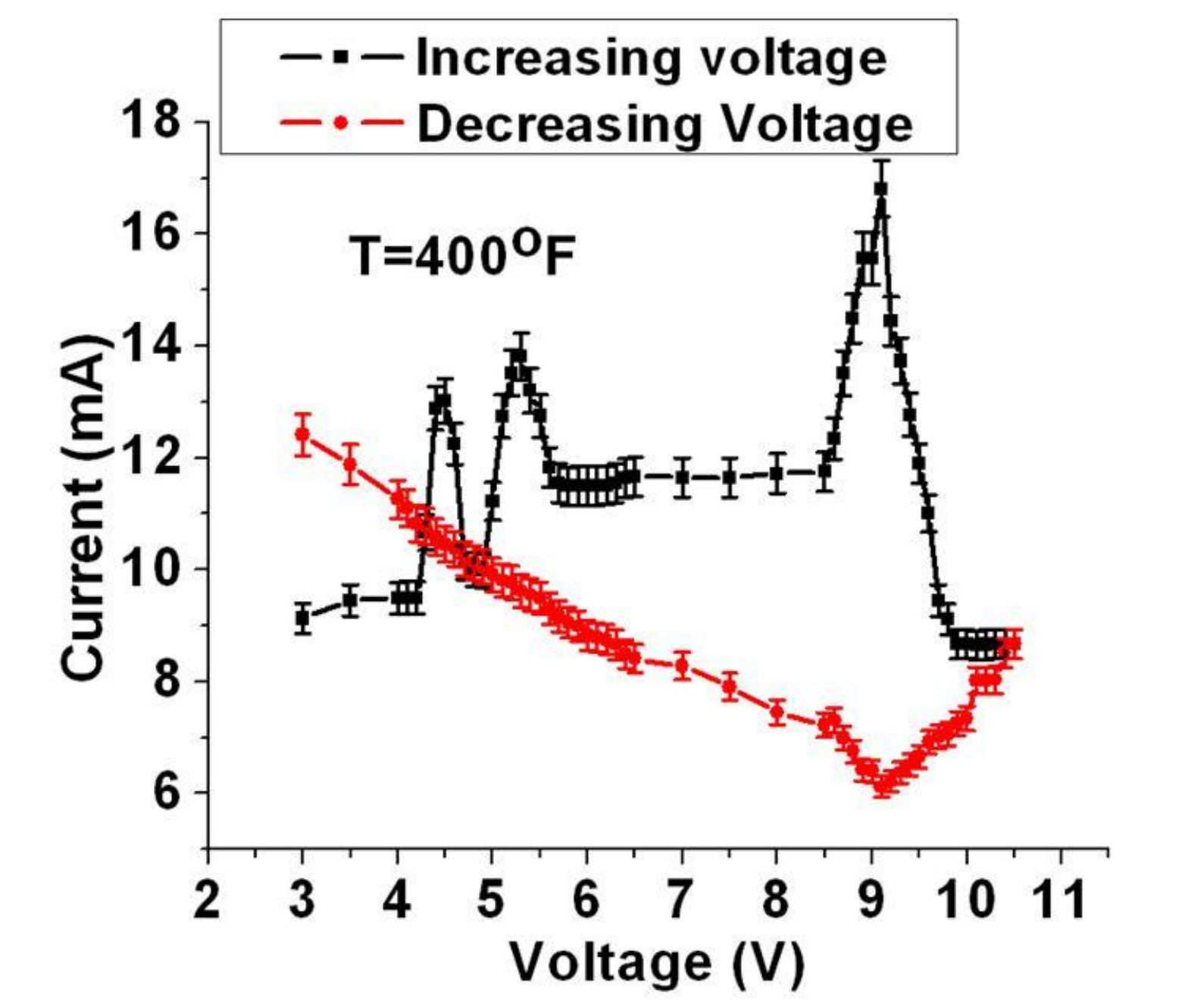}
\caption{I-V characteristics of colemanite at $400^{0}$F. The output current was 0.1 mA when all electronics were turned off with just the oven stabilized to $400^{0}$F. This offset current is the source of error in the experiment.}
\end{figure}

\subsection{\label{sec:level2}I-V characteristics at a high temperature}

Figure 3 shows the I-V characteristics of colemanite at a temperature of $400^{0}$F. The I-V plot exhibits hysteresis at a high temperature. As the voltage is increased from 3 V, the current is almost stable at 9.5 V, until the voltage reaches 4.2 V. It then exhibits two sharp peaks at 4.5 V and 5.2 V. After that, the current is again almost stable at about 11.8 mA between voltages 5.5 V and 8.4 V. It then exhibits a sharp peak at 8.8 V. After 9.3 V the current stabilized at 8.5 mA.  The appearance of the peaks can be explained by Frenkel-Poole emissions~\cite{Pintilie} as follows. Different defects form intraband donor or acceptor levels which act as traps for electrons. At high temperatures, these electrons gain enough energy to get out of their localized state and move to the conduction band. Since these transitions take place between quantum energy levels, they occur at specific energies. This is the reason why the peaks appear only at certain voltages. The water molecules in colemanite also play a significant role in the process of conduction. At a high temperature, the water molecules undergo 180 degree flips about the line bisecting the H--O--H angle~\cite{Watton}. This flip causes redistribution of protons over the energy levels of the OH---H bonds. This creates a few free protons which contributes to the conduction mechanism by hopping~\cite{Gavrilova}.

As the voltage is decreased, the current decreases almost in a straight line and exhibits ohmic behavior until the voltage reaches 9V, as shown in Fig. 3. As the voltage is further decreased, the current steadily increases until the voltage reaches 3 V. The crystal offers negative resistance between 9V and 3 V. This behavior is similar to the I-V characteristics at room temperature. However, the transition from ohmic resistance to a negative resistance is observed at a higher voltage, at $400^{0}$F than at room temperature. This is because, the high temperature causes the electrons to easily break free from the traps and start conducting. Thus, the valley appears sooner in Fig. 3 than in Fig. 2. The presence of these defects create anharmonicity in the crystal lattice. The anharmonic rearrangement of the crystal structure at a high temperature is the cause of pyroelectricity in colemanite. This has been observed in ferroelectric Lithium niobate crystals \cite{Peng}.

\subsection{\label{sec:level2}I-V characteristics under the influence of near UV light}

Figure 4 shows the I-V characteristics of colemanite when excited by 395 nm near UV light. The plot exhibits hysteresis. As the voltage increases, the current increases almost in a straight line, as shown in Fig.~4. The current is constant between 9 V to 12 V. The constant current is due to the alignment of internal dipole moments in a single direction. When the voltage is decreased, the current steadily decreases in a straight line. The crystal is ohmic in both directions of changing the voltage. This can be explained by the fact that colemanite is a polycrystalline material. So each grain boundary is photovoltaic. Since the grains add in series, the crystal behaves ohmic. Moreover, the ferroelectric domains in colemanite are photovoltaic, with each domain wall acting as a contact connecting the neighboring photovoltaics. The domain walls too add in series, contributing to the ohmic resistance. The additive photovoltaic microstructures such as grain boundaries, domains and others create a large photovoltage. The higher the photovoltaic microstructure, more is the photovoltage generated. Hence, colemanite could be a replacement for the present day semiconductors in photovoltaic devices that generate low voltages.

\begin{figure}
\centering\includegraphics [width=3.5in,height=2.8in]{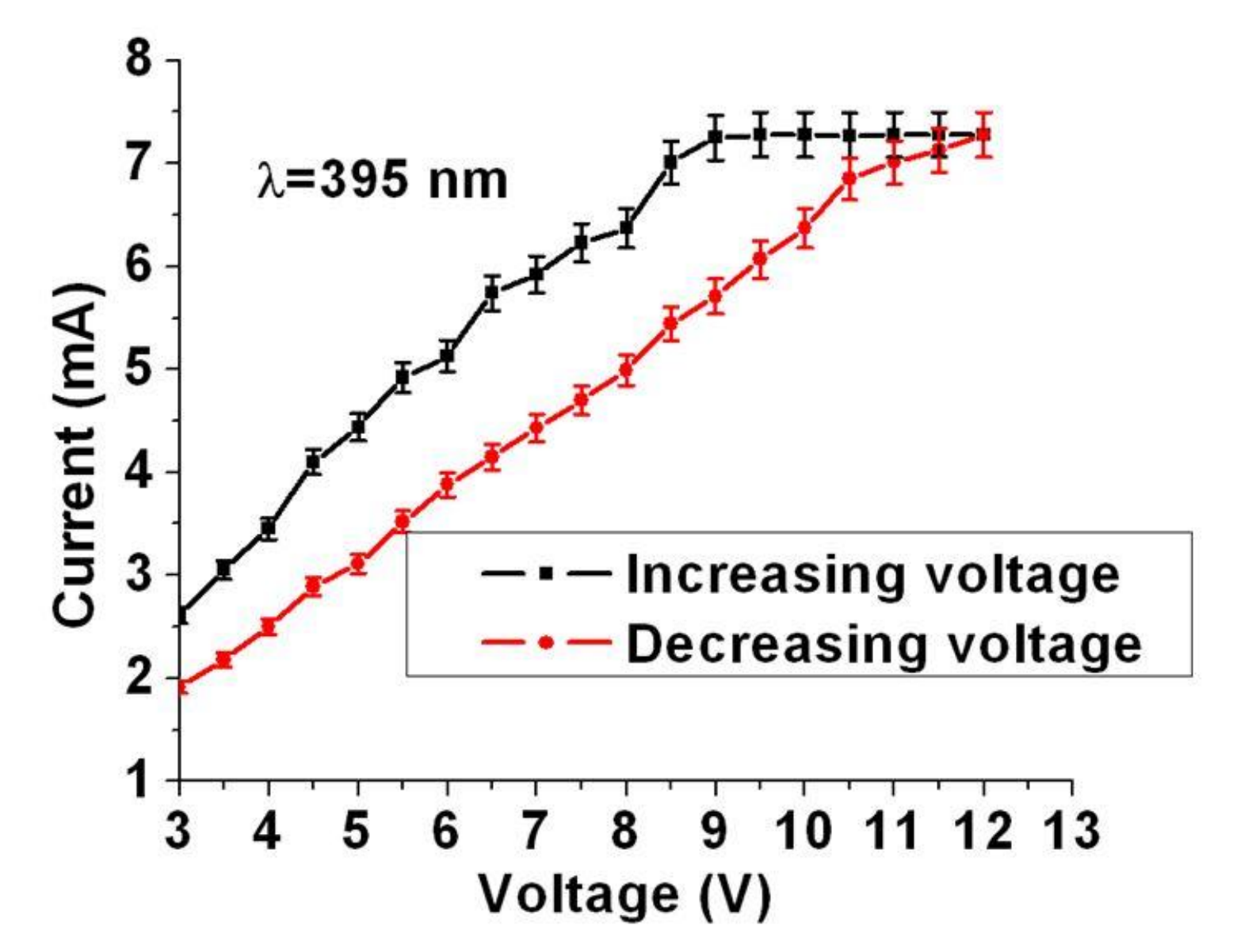}
\caption{I-V characteristics of colemanite under excitation by near UV light of wavelength 395 nm. When the light source is turned on, but the power supply is turned off, a small current of 0.1 mA appears after 2 seconds. This is due to the creation of light-induced electron-hole pairs.  }
\end{figure}

\section{\label{sec:level5}CONCLUSIONS}

We investigated the I-V characteristics of ferroelectric colemanite at room temperature, at a high temperature of $400^{0}$~F, and under the influence of near UV light and observed that hysteresis effects are present in the I-V characteristics in all three cases. At the room temperature, the I-V plot shows a change from ohmic resistance to a negative resistance, when the voltage is increased as well as decreased. This happens due to the presence of the defects in the crystal. We argued that at lower voltages, the defects act as low resistance conducting paths between the electrodes, which increases the current. At higher voltages, the conducting paths break down and the defects act as traps for free carriers decreasing the current. This negative resistance exhibited by colemanite in our experiment can also be used in negative resistance devices such as in amplifiers and microwave energy generators~\cite{golio}. The sharp peaks at the higher temperature is explained by conduction mechanisms in colemanite such as Frenkel-Poole emissions. Another conduction mechanism leading to such peaks is due to the hopping of free protons~\cite{Gavrilova} that are created by 180 degree flips of the water molecules at high temperatures~\cite{Watton}.

We also present our results when colemanite is excited with near UV light. The excitation reveals ohmic resistance when the voltage is changed in both directions, that is increased as well as decreased. This new behavior observed in the I-V characteristics of colemanite can be explained by the presence of photovoltaic microstructures such as grain boundaries, ferroelectric domains and others. These results show colemanite can be used to generate high photovoltages, while conventional semiconductors can only generate low photovoltages limiting their applications in many devices. This opens up exciting possibilities of replacing conventional semiconductors with colemanite in photovoltaic devices.

\end{document}